\documentclass[prl,aps,twocolumn,nofootinbib,superscriptaddress]{revtex4-1} 

\usepackage{pdfpages}
\makeatletter
\AtBeginDocument{\let\LS@rot\@undefined}
\makeatother

\usepackage{amsmath,verbatim}
\usepackage{bbm}
\usepackage{graphicx}
\usepackage{color}
\usepackage{amssymb}
\usepackage{amsfonts}
\usepackage{tikz}
\usepackage{xparse}
\usepackage{ifthen}
\usepackage{lipsum}
\usepackage{units}

\usepackage[normalem]{ulem}
\usepackage[colorlinks=true,linkcolor=blue,citecolor=blue,urlcolor=blue]{hyperref}
\usepackage{siunitx}
\usepackage{dsfont}
\begin{document}

\newcommand{\tr}{\mathop{\mathrm{tr}}}
\renewcommand{\geq}{\geqslant}
\renewcommand{\leq}{\leqslant}

\title{Chiral Ising Gross-Neveu criticality of a single Dirac cone: A quantum Monte Carlo study}

\author{S. Mojtaba Tabatabaei}
\thanks{These two authors contributed equally.}
\affiliation{Department of Physics, Sharif University of Technology, Tehran 14588-89694, Iran}

\author{Amir-Reza Negari}
\thanks{These two authors contributed equally.}
\affiliation{Department of Physics, Sharif University of Technology, Tehran 14588-89694, Iran}

\author{Joseph Maciejko}
\affiliation{Department of Physics \& Theoretical Physics Institute (TPI), University of Alberta, Edmonton, Alberta T6G 2E1, Canada}

\author{Abolhassan Vaezi}
\email{Corresponding Author: vaezi@sharif.edu}
\affiliation{Department of Physics, Sharif University of Technology, Tehran 14588-89694, Iran}

\begin{abstract}
We perform large-scale quantum Monte Carlo simulations of SLAC fermions on a two-dimensional square lattice at half filling with a single Dirac cone with $N=2$ spinor components and repulsive on-site interactions. Despite the presence of a sign problem, we accurately identify the critical interaction strength $U_c = 7.28 \pm 0.02$ in units of the hopping amplitude, for a continuous quantum phase transition between a paramagnetic Dirac semimetal and a ferromagnetic insulator. Using finite-size scaling, we extract the critical exponents for the corresponding $N=2$ chiral Ising Gross-Neveu universality class: the inverse correlation length exponent $\nu^{-1} = 1.19 \pm 0.03$, the order parameter anomalous dimension $\eta_{\phi} = 0.31 \pm 0.01$, and the fermion anomalous dimension $\eta_{\psi} = 0.136 \pm 0.005$.
\end{abstract}
\maketitle

{\it Introduction.}---Massless Dirac fermions have been identified as the relevant low-energy quasiparticles in various condensed matter systems including graphene, topological insulators, $d$-wave superconductors, Weyl semimetals, and ultracold fermions in optical lattices~\citep{RevModPhys.78.373,RevModPhys.81.109,RevModPhys.82.3045,RevModPhys.83.1057,PhysRevB.83.205101,PhysRevLett.111.185307,doi:10.1080/00018732.2014.927109}. Nonetheless, strong interactions can generate a finite mass for the Dirac fermions and spontaneously break some of the symmetries of the model. The quantum phase transitions at which this occurs are typically described by the Gross-Neveu (GN) university classes~\cite{[{For a recent review of quantum critical phenomena in Dirac systems, see }]boyack2021}. In particular, a single Dirac cone in (2+1)D subject to on-site repulsive interactions---such as can be found on the surface of a correlated topological insulator---can develop an Ising-type ferromagnetic (FM) order, which generates a $\mathbb{Z}_2$ symmetry-breaking FM mass gap~\cite{xu2010,neupert2015}. For a chemical potential at the Dirac point, the quantum critical point (QCP) of the resulting transition from semimetal (SM) to insulator is believed to belong to the chiral Ising GN universality class~\cite{PhysRevD.10.3235,zinn-justin1991,Rosenstein1993,PhysRevB.96.165133,PhysRevD.96.096010,PhysRevB.98.125109} with $N=2$ Dirac spinor components.

Useful insights for the $N=2$ chiral Ising GN universality class have been obtained from several approaches including the conformal bootstrap, the functional renormalization group (fRG), and analytical field theory methods such as large-$N$ and $\epsilon$ expansions. However, these methods so far yield inconsistent results. For example, while the conformal boostrap~\citep{Iliesiu2018} predicts $\nu^{-1}=0.86$, fRG~\citep{PhysRevD.91.125003} and the $\epsilon$ expansion~\citep{PhysRevB.98.125109} predict $\nu^{-1}=1.229$ and $\nu^{-1}=1.276$, respectively. (For other critical exponents, see Table~\ref{table1}.) These significant discrepancies demand a resolution from numerically exact quantum Monte Carlo (QMC) simulations which have been unavailable thus far. The lack of QMC studies of this problem originates in part from fermion-doubling theorems which state that a local lattice model cannot realize a single symmetry-protected Dirac cone~\cite{huang2020}. Indeed, all previous QMC studies of chiral Ising GN criticality have utilized local lattice models and thus could only access even numbers of Dirac cones, e.g., $N=4$~\cite{Wang2014,Li2015,PhysRevB.93.155157,huffman2020} and $N=8$~\cite{PhysRevB.97.081110,PhysRevLett.122.077601,zhang2019,PhysRevB.101.064308,zhang2020}.

\begin{figure}[t]
\includegraphics[width=\columnwidth]{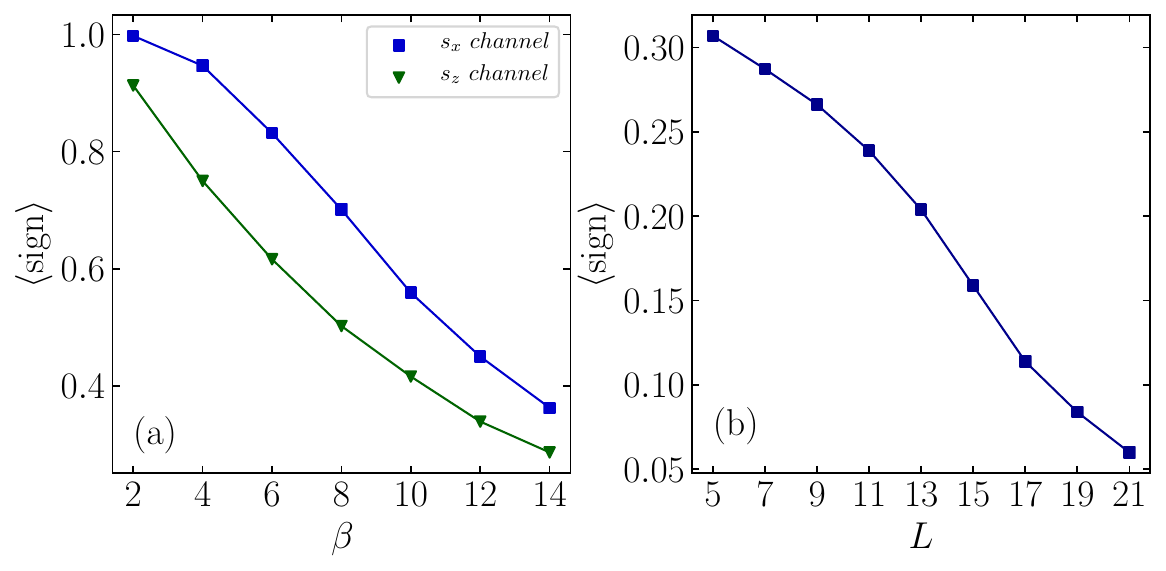}
\caption{\label{fig:sgn} Behavior of the sign problem in PQMC with $g_\text{GW} = 0.17U$ close to the QCP ($U=7.275$). (a) Decoupling the Hubbard interaction in the $s_x$ channel enhances the average sign compared to the usual $s_z$ channel (here $\beta\equiv 2\Theta$). (b) Average sign of PQMC at $2\Theta = 14$ ($\beta_{\rm eff} \approx 25 \pm 1$) and $U=7.275$ as a function of linear system size $L$.}
\end{figure}

In this paper, we instead use a nonlocal lattice realization of a single Dirac fermion with $N=2$ spinor components, known as the SLAC fermion~\cite{drell1976,Li2018,PhysRevLett.123.137602}, subject to an on-site Hubbard repulsion. By employing a state-of-the-art auxiliary-field QMC algorithm, we identify and investigate its FM QCP for the first time~\cite{SM}. The model is not entirely sign-problem free, but the sign problem is benign at the QCP (Fig.~\ref{fig:sgn})~\cite{CM1}. 
In this work, we have have taken up to several billion measurements to keep the statistical error below $0.2\%$~\cite{CM2}. 
This approach allows us to circumvent the sign problem and accurately extract the critical exponents of the $N=2$ chiral Ising GN universality class (Table~\ref{table1}), our main result.

\begin{table}
\begin{centering}
\begin{tabular}{lccc}
\hline 
 & $\nu^{-1}$ & $\eta_{\phi}$ & $\eta_{\psi}$\tabularnewline
\hline 
\hline 
this work (QMC) & $1.19\pm0.03$ &$0.31\pm0.01$&$0.136\pm0.005$ \tabularnewline
conf. bootstrap~\citep{Iliesiu2018} & $0.86$ & $0.320$ & $0.134$\tabularnewline
fRG~\citep{PhysRevD.91.125003} & $1.229$ & $0.372$ & $0.131$\tabularnewline
$\epsilon$ expansion~\citep{PhysRevB.98.125109} & $1.276$ & $0.2934$ & $0.1400$\tabularnewline
\hline 
\end{tabular}\caption{\label{table1}Our QMC evaluation of the critical exponents for the $N=2$ chiral Ising GN universality class, compared with previous estimates.}
\par\end{centering}
\end{table}

{\it Model.}---We consider an $L\times L$ square lattice with unit lattice constant having a single linearly dispersing Dirac cone in its first Brillouin zone. The free Hamiltonian in momentum space is given by:
\begin{equation}
H_{0}=\sum_{{\bf p}}\Psi_{{\bf p}}^{\dagger}(p_{x}\sigma_{x}+p_{y}\sigma_{y})\Psi_{{\bf p}},
\end{equation}
with $\Psi_{{\bf p}}^{\dagger}=(c_{{\bf p}\uparrow}^{\dagger},c_{{\bf p}\downarrow}^{\dagger})$ where $c_{{\bf p}\sigma}^{(\dagger)}$ is the electron annihilation (creation) operator with momentum ${\bf p}=(p_{x},p_{y})$ and spin $\sigma$, and $\sigma_{\alpha}$, $\alpha=x,y,z$ are the Pauli matrices operating on the spin degree of freedom. We extract the real-space representation of the above Hamiltonian by performing a Fourier transformation, which yields:
\begin{equation}
H_{0}=\sum_{\bf i}\sum_{{\bf R}}\left(t_{{\bf R}}c_{\bf i\uparrow}^{\dagger}c_{\bf i+ R\downarrow}+\mathrm{h.c.}\right),
\end{equation}
where $c_{\bf i\sigma}^{(\dagger)}$ is the electron annihilation (creation) operator on site $
\bf i$ with spin $\sigma$, and $t_{{\bf R}}$ denotes the electron hopping amplitude between site $\bf i$ and $\bf i+R$. Here
${\bf R}=(R_x,R_y)$ enumerates all neighbors of site $\bf i$ along the $x$ and $y$ directions. The explicit form of $t_{{\bf R}}$ is: 
\begin{equation}\label{tR}
t_{{\bf R}}=\frac{i(-1)^{R_{x}}}{\frac{L}{\pi}\sin(\frac{\pi R_{x}}{L})}\delta_{R_{y},0}+\frac{(-1)^{R_{y}}}{\frac{L}{\pi}\sin(\frac{\pi R_{y}}{L})}\delta_{R_{x},0},
\end{equation}
where the overall hopping amplitude has been set to unity. Note that Eq.~(\ref{tR}) introduces electron hopping beyond nearest neighbors. We add a local repulsive Hubbard interaction, 
\begin{equation}
H_{U}=U\sum_{\bf i}\left(n_{\bf i\uparrow}-1/2\right)\left(n_{\bf i\downarrow}-1/2\right),
\end{equation}
where $U>0$ is the interaction strength and $n_{\bf i\sigma}=c_{\bf i\sigma}^{\dagger}c_{\bf i\sigma}$ is the electron number operator. For sufficiently large $U$, we expect long-range Ising FM order in the $z$ direction, which breaks time-reversal symmetry spontaneously and gaps out the Dirac cone. At half-filling, the single-particle density of states vanishes, thus we expect a line of finite-temperature transitions that terminates at a zero-temperature QCP with finite critical interaction strength $U_c$~\cite{PhysRevB.93.155157}.

{\it QMC method.---}We employ a projector QMC (PQMC) method to analyze the quantum phase transition in our model system. In this method, the ground-state expectation value of an observable $O$ is calculated using imaginary-time propagation of a trial wave function
$\left|\Psi_{T}\right\rangle $ via $\frac{\left\langle \Psi_{0}\left|O\right|\Psi_{0}\right\rangle }{\left\langle \Psi_{0}|\Psi_{0}\right\rangle }=\lim_{\Theta\rightarrow\infty}\frac{\left\langle \Psi_{T}\left|e^{-\Theta H}Oe^{-\Theta H}\right|\Psi_{T}\right\rangle }{\left\langle \Psi_{T}\left|e^{-2\Theta H}\right|\Psi_{T}\right\rangle }$.
Here, we follow the approach introduced in Ref.~\citep{vaezi2018unified} and choose an interacting trial wave function to further enhance the performance and convergence of the PQMC algorithm. We consider a Gutzwiller-projected wave function  $\left|\Psi_{T}\right\rangle =e^{-g_\text{GW}\sum_{\bf i}n_{{\bf i},\uparrow}n_{{\bf i},\downarrow}}\left|{\rm FS}\right\rangle $ which can be easily implemented as our trial state within QMC. Here, $\left|{\rm FS}\right\rangle $ is the noninteracting Fermi sea, and $g_\text{GW}$ is the Gutzwiller variational parameter whose optimal value follows the approximate relation $g_\text{GW}\approx 0.17 U$. We perform calculations for system sizes $L\in\{5,7,9,11,13,15,17,19\}$, and use an imaginary projection time of $2\Theta=14$ which is long enough to obtain ground-state properties. To demonstrate this, in the Supplemental Material \cite{SM} we compare the performance of PQMC with a nontrivial Gutzwiller-projected state against regular PQMC ($g_\text{GW}=0$) as well as finite-temperature QMC. We find that the algorithm with $g_\text{GW}\neq 0$ converges to the ground state the fastest. Moreover, an effective inverse temperature $\beta_{\rm eff}$ can be defined for a given projection time $\Theta$ such that the PQMC results are approximately equivalent to finite-temperature QMC results at temperature $T=1/\beta_{\rm eff}$. We find that $\beta_{\rm eff}\left(g_\text{GW} \neq 0\right) \approx 2\Theta + 11 \pm 1$ while $\beta_{\rm eff}\left(g_\text{GW}=0\right) \approx 2\Theta + 7 \pm 1$. Our PQMC method with $g_{\rm GW}\neq 0$ and $2\Theta=14$ thus allows us to effectively reach temperatures as low as $\beta_{\rm eff} = 25 \pm 1$, which is sufficient to elucidate ground-state physics.

Although QMC is an unbiased method and is very effective for studying lattice models of strongly correlated electrons, its negative sign problem hinders its application to many problems of interest~\cite{troyer2005}. Nonetheless, the sign problem in QMC depends highly on the model's formulation, such that one may improve the energy scales that QMC can reach by choosing appropriately the Hubbard-Stratonovich (HS) decoupling of the interaction term. For the present model, the average sign is significantly higher if we decouple the interaction in the $s_x$ or $s_y$ channels rather than the usual $s_z$ channel [Fig.~\ref{fig:sgn}(a)]~\cite{SM}. With that decoupling, Fig.~\ref{fig:sgn}(b) shows that the average sign of our model at the QCP is not very severe, and we can reach sufficiently low temperatures to accurately predict ground-state properties.

{\it FM transition.}---We probe FM ordering in our model by computing the spin-spin correlation function, 
\begin{equation}\label{Mij}
M_{\bf ij}=\bigl\langle s_{z,\bf i}s_{z,\bf j}\bigr\rangle,
\end{equation}
whose Fourier transform is the spin structure factor: 
\begin{equation}
S({\bf k})=\frac{1}{L^4}\sum_{\bf ij}e^{i{\bf k}\cdot({\bf i}-{\bf j})}M_{\bf ij},
\end{equation}
where $s_{z,\bf i}=\frac{1}{2}\left(n_{\bf i\uparrow}-n_{\bf i\downarrow}\right)$
denotes the $z$ component of the electron spin operator at site $\bf i$. In the broken-symmetry phase at large $U$, we expect long-range order at wave vector ${\bf k}=0$ and the condensation of the $s_{z,\bf i}$ operator in the thermodynamic limit. 

\begin{figure}[t]
\includegraphics[width=\columnwidth]{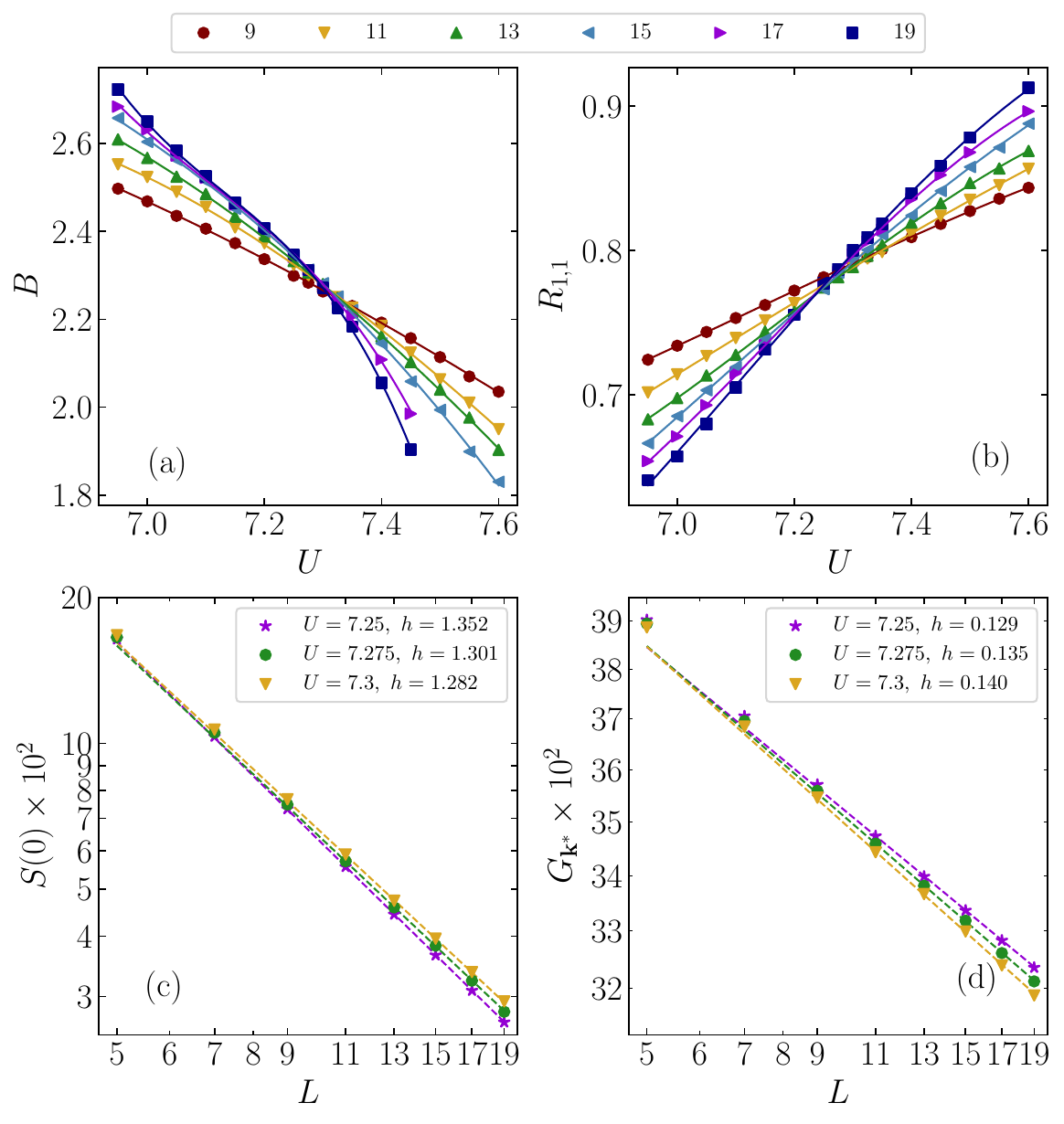}
\caption{\label{fig:binder} (a) Binder ratio $B$ and (b) correlation ratio $R_{1,1}$ as a function of $U$ for various $L$ (various symbols). The crossing point corresponds to $U_{c}$. We identify $7.25 <U_c \lesssim 7.3$ using these two methods. (c) FM spin susceptibility $S_{{\bf k}=0}$ and (d) equal-time fermion Green's function $G_{{\bf k}={\bf k}^*}$ for various system sizes close to the critical point ($U=7.25,7.275,7.3$). The observed linear behavior on a log-log scale is consistent with the expected power-law decay $S_{{\bf k}=0} \sim L^{-(1+\eta_{\phi})}$ and $G_{{\bf k}={\bf k}^*} \sim L^{-\eta_{\psi}}$ at criticality. The negative of the slope $h$ is included for each $U$. We find $\eta_{\phi}=h-1 \approx 0.31$ and $\eta_{\psi}=h \approx 0.135$ by taking the average across all three values of $U$.}
\end{figure}

\begin{figure}[t]
\includegraphics[width=\columnwidth]{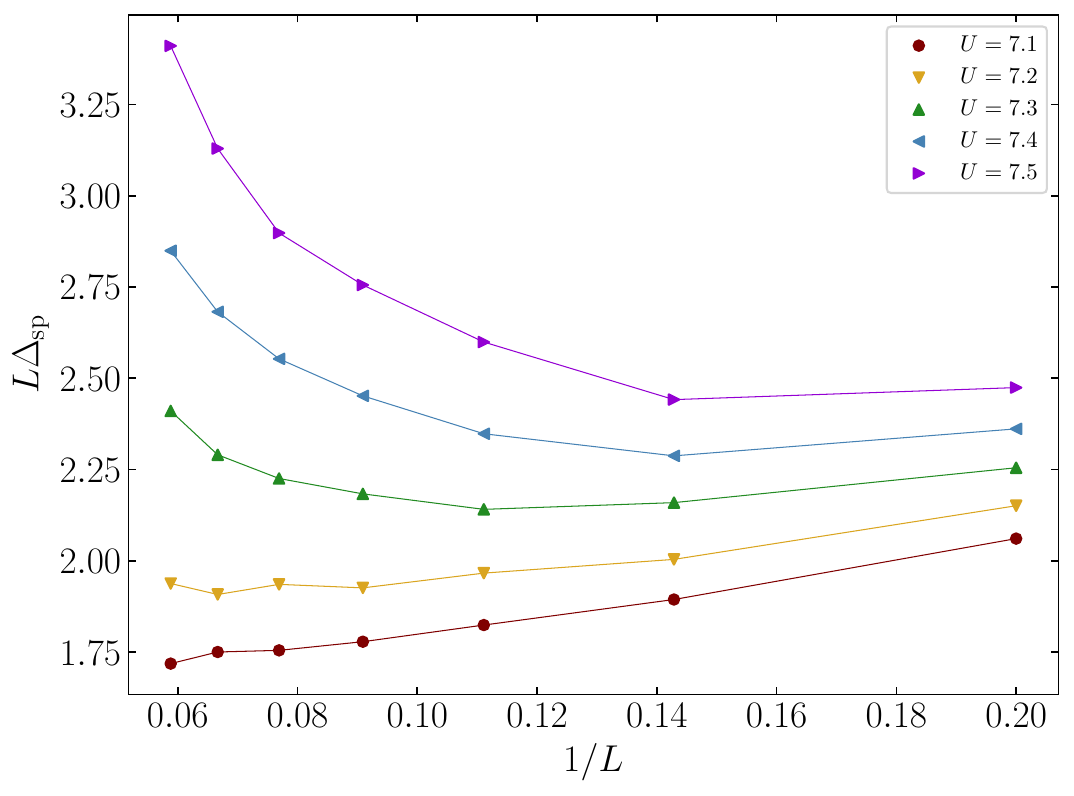}
\caption{\label{fig:gap}Fermion single-particle gap $\Delta_\text{sp}$ as a function of $1/L$ for various values of $U$, which suggests $7.2<U_c<7.3$.}
\end{figure}

To explore the SM-to-FM QCP in QMC, we use two dimensionless quantities: the Binder ratio, defined here as
\begin{equation}
B \equiv\frac{\sum_{\bf ijkl}\bigl\langle s_{z,\bf i}s_{z,\bf j}s_{z,\bf k}s_{z,\bf l}\bigr\rangle}{\left(\sum_{\bf ij}\bigl\langle s_{z,\bf i}s_{z,\bf j}\bigr\rangle\right)^{2}},
\end{equation}
and the correlation ratio, defined as:
\begin{equation}
R_{1,1}\equiv 1-\frac{S({\bf k}={\bf k}^*)}{S({\bf k}=0)},
\end{equation}
where we define ${\bf k}^*\equiv\frac{2\pi}{L}(\hat{x}+\hat{y})$. Long-range FM ordering makes $S({\bf k}=0)$ diverge and hence implies $R_{1,1}\rightarrow1$ in the thermodynamic limit $L\rightarrow\infty$. In the disordered SM phase, the correlation ratio vanishes in the thermodynamic limit since $S\left({\bf k}\rightarrow 0 \right) \to S\left({\bf k}=0\right)$. At the QCP, both $B$ and $R_{1,1}$ are independent of $L$ up to finite-size corrections. Therefore, we pinpoint the QCP by plotting these ratios as a function of $U$ for various lattice sizes, and look for a crossing point of the curves. Using the Binder ratio, we identify the QCP to be $7.275 \leq U_c \leq 7.3$ [Fig.~\ref{fig:binder}(a)]. The correlation ratio suggests the compatible result $7.25 \leq U_c \leq 7.275$ [Fig.~\ref{fig:binder}(b)].

To further corroborate these results, we also measure the fermion excitation gap $\Delta_\text{sp}$ as a function of $L$ and $U$ using the unequal-time fermion Green's function~\cite{SM}. 
In the thermodynamic limit, we expect $L\Delta_\text{sp}\left(L\to \infty,U < U_c\right)\to 0$ in the gapless SM phase. Thus, we can estimate the position of the QCP by plotting $L\Delta_\text{sp}\left(L,U\right)$ against $1/L$ and extrapolating to $L = \infty$ (see Fig.~\ref{fig:gap}). This suggests $7.2<U_c<7.3$, consistent with the previous two approaches. These three methods combined indicate that $U_c \approx 7.275$. In the Supplemental Material \cite{SM}, we have computed $B$ and $R_{1,1}$ using finite-temperature QMC with $\beta = L$~\citep{liu2019superconductivity,PhysRevLett.122.077601,liu2021exotic} for $L$ up to $15$ and achieve $7.25<U_c<7.3$, consistent with our PQMC results.

{\it Critical exponents.}---Having obtained a good estimate of $U_c$, we now turn to calculating universal critical exponents directly at the QCP. Those exponents describe the power-law decay of various correlation functions at the QCP. In Fig.~\ref{fig:binder}(c), we plot the FM spin susceptibility, $S({\bf k}=0)$, for interaction strengths $U=7.25, 7.275$, and $7.3$. The spin susceptibility is expected to decay as $L^{-(1+\eta_{\phi})}$ at the critical point for an $L\times L$ system. Figure~\ref{fig:binder}(c) shows that the finite-size effects in the two-particle spin (bosonic) sector are insignificant as all data points follow a single straight line on a log-log scale. Our results in Fig.~\ref{fig:binder}(c) thus suggest the anomalous dimension of the bosonic order parameter, $\eta_{\phi}$, satisfies $0.282<\eta_\phi<0.352$. Likewise, the equal-time fermion single-particle Green's function in momentum space, $G_{{\bf k}={\bf k}^*}$, must decay as $L^{-\eta_{\psi}}$, where $\eta_\psi$ is the anomalous dimension of the fermion operator at criticality. Accordingly, Fig.~\ref{fig:binder}(d) shows that $0.129<\eta_{\psi}<0.140$. We obtained these numbers by taking the last five data points ($L=11,13,15,17,19$) for fermions. We see that $L=9$ follows the same line while $L=5,7$ exhibit visible deviations. This implies that finite-size effects are more pronounced in the fermionic sector. Among the three interaction strengths used in Fig.~\ref{fig:binder}(c-d), our Binder/correlation ratio analysis suggests $U_c$ is closer to $7.275$. Thus we conclude $\eta_{\phi} \approx 0.30 \pm 0.02$ and $\eta_{\psi} \approx 0.135 \pm 0.005$. 

\begin{figure}[t]
\includegraphics[width=\columnwidth]{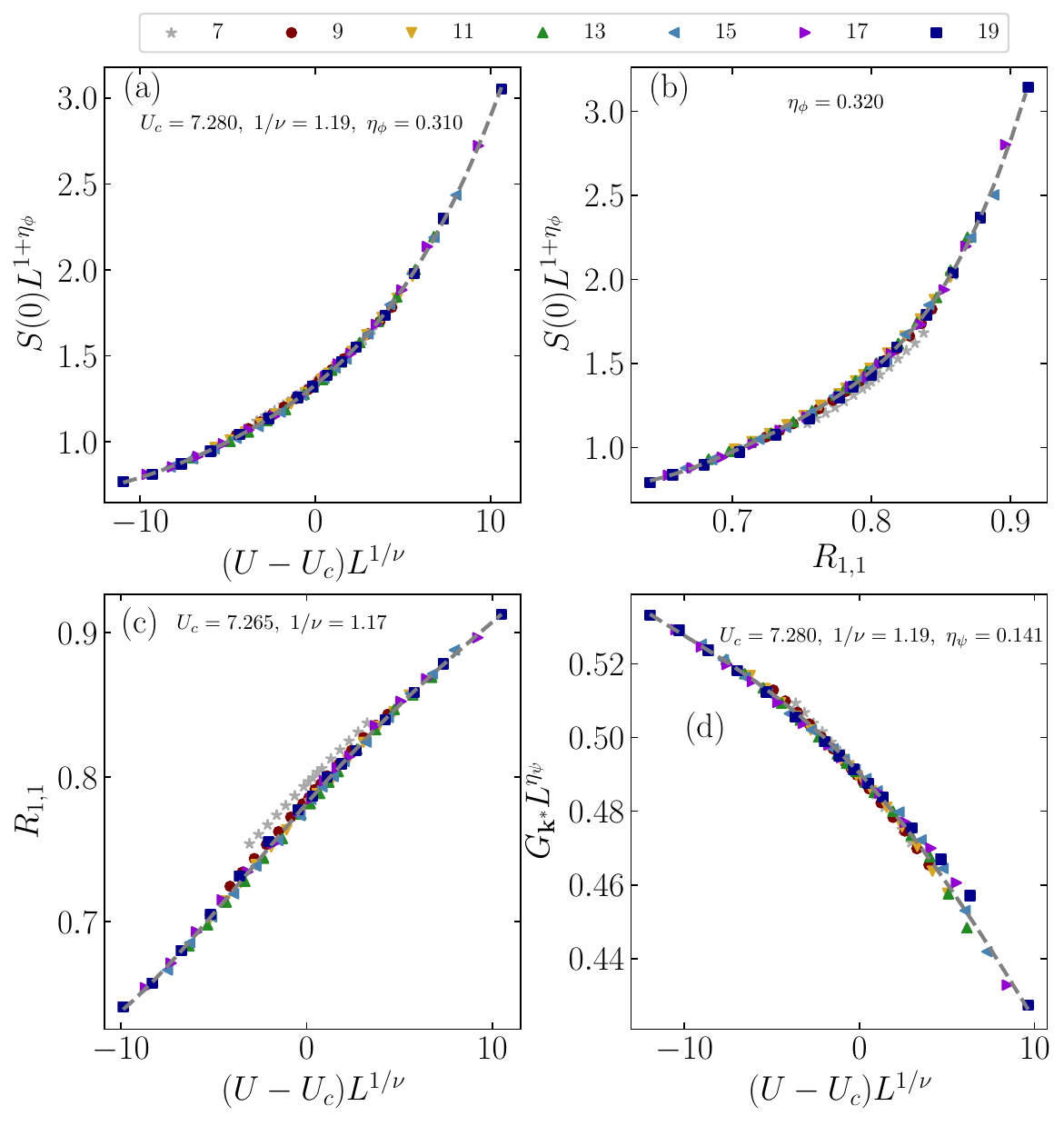}
\caption{\label{fig:f3} Data collapse using the leading-order scaling hypothesis to estimate $U_c$ and the critical exponents $\nu^{-1}$, $\eta_\phi$, and $\eta_\psi$. The results are based on system sizes  $9\leq L\leq 19$, although we have plotted $L=7$ using the estimated critical exponents as well.}
\end{figure}

Alternatively, we can use the scaling hypothesis and data collapse near (but away from) the QCP to simultaneously obtain estimates of the critical exponents as well as $U_c$. Scaling forms for bosonic and fermionic correlation functions can be used to extract $\eta_\phi$ and $\eta_\psi$. We begin with the spin structure factor. At $\beta = \infty$ or $\beta = L$ and near the QCP, scaling analysis reveals that~\cite{PhysRevX.6.011029}:
\begin{equation}\label{scalingL}
L^{1+\eta_{\phi}}S_{{\bf k} = 0} (L,U) = \left(1+\alpha_1 L^{-\omega_1}\right) f_1(uL^{1/\nu}),
\end{equation}
where $u = U-U_c$, $\nu$ is the correlation length exponent, and $f_1$ is a smooth scaling function of $uL^{1/\nu}$. The term proportional to $L^{-\omega_1}$ is an effective correction-to-scaling term which can be ignored for large systems. For $S_{{\bf k} = 0}$ we find that those corrections are negligible and we achieve satisfactory results by keeping the leading scaling term. Such a simplified scaling hypothesis, namely $L^{1+\eta_{\phi}}S_{{\bf k} = 0} (L,U) = f_1\left(\left(U-U_c\right)L^{1/\nu}\right)$, allows the following data-collapse method to extract the critical exponents. By plotting all available data points in the $L^{1+\eta_{\phi}}S_{{\bf k} = 0} (L,U)$ combination against $(U-U_c)L^{1/\nu}$ and tuning $U_c$, $\nu$, and $\eta_{\phi}$ to achieve a single smooth curve rather than scattered data points, we can identify both the critical exponents $\nu$ and $\eta_\phi$ and the critical point $U_c$ [Fig.~\ref{fig:f3}(a)]. This method yields $U_c\approx 7.280$, $\nu^{-1}\approx 1.19$, and $\eta_\phi\approx 0.310$. Again, finite-size effects are minimal here: we see in Fig.~\ref{fig:f3}(a) that data points for systems as small as $L=7$ also collapse to the fitting curve.

Additionally, near the QCP, the correlation ratio $R_{1,1}$ behaves as a universal function of $(U-U_{c})L^{1/\nu}$ and $L^{z}/\beta$ where $z$ is the dynamical critical exponent and $\beta$ the inverse temperature. Here emergent Lorentz symmetry at the QCP implies $z=1$. In Fig.~\ref{fig:f3}(c), data collapse of $R_{1,1}$ yields the estimates $U_c\approx 7.265$ and $\nu^{-1}\approx 1.17$. We can also plot $L^{1+\eta_{\phi}}S({\bf k} = 0)$ against $R_{1,1}$ to extract $\eta_{\phi}\approx 0.320$ [Fig.~\ref{fig:f3}(b)]. The main advantage of this method compared to that used in Fig.~\ref{fig:f3}(a) is that neither $U_c$ nor $\nu$ need to be determined.

Similarly, to compute the fermion anomalous dimension $\eta_{\psi}$, we can utilize the following scaling hypothesis in the proximity of the QCP:
\begin{equation}\label{scalingG}
L^{\eta_{\psi}}G_{{\bf k}={\bf k}^*} (L,U) = \left(1+\alpha_2 L^{-\omega_2}\right) f_2(u L^{1/\nu}),  
\end{equation}
where:
\begin{equation}
G_{{\bf k}={\bf k}^*}(L,U) \equiv  \frac{1}{L^4}\sum_{\bf ij} e^{i\bf k^*\cdot\left(i-j\right)}\langle c_{\bf i\uparrow}^\dag c_{\bf j\downarrow}\rangle,
\end{equation} 
and $f_2$ is another smooth scaling function. Applying data collapse to $G_{{\bf k^*}} (L,U)$ yields satisfactory results, especially for $L \geq 9$ [Fig.~\ref{fig:f3}(d)]. We find $\eta_{\psi} \approx 0.141$, $U_c \approx 7.280$, and $\nu^{-1} \approx 1.19$.

Combining our results directly obtained at the QCP and those extracted from data collapse in the vicinity of the QCP, we obtain a consistent set of critical exponent estimates with error bars that reflect the totality of our results (Table~\ref{table1}). In the Supplemental Material \cite{SM}, we have investigated the impact of corrections to scaling on the critical exponents we extract. Although the quality of data collapse increases significantly upon introducing the associated free parameters $\alpha_{1,2}$ and $\omega_{1,2}$ in Eqs.~(\ref{scalingL}-\ref{scalingG}), we find that the exponent values remain unchanged within the statistical error bar.

{\it Summary and outlook.}---In summary, we applied a PQMC method with Gutzwiller-projected trial state to study the quantum phase transition from paramagnetic Dirac semimetal to ferromagnetic insulator in a model of a single two-component Dirac fermion in (2+1)D subject to an on-site repulsive Hubbard interaction $U$. We also performed finite-temperature QMC calculations for the same model. Both methods yield consistent results, from which we conclude that the phase transition is continuous and happens at $U_c= 7.28 \pm 0.02$ in units of the fermion hopping amplitude. Besides determining the position of the QCP, our main result is a numerically exact determination of the critical exponents of the associated $N=2$ chiral Ising GN universality class: the inverse correlation length exponent $\nu^{-1}= 1.19 \pm 0.03$, the order parameter anomalous dimension $\eta_{\phi}= 0.31 \pm 0.01$, and the fermion anomalous dimension $\eta_{\psi} = 0.136 \pm 0.005$.

The discrepancy between the conformal bootstrap and the other methods in Table~I for $\nu^{-1}$ is more significant than for other critical exponents. Interestingly, this appears to be common to other Ising GN universality classes (see Table IV in Ref.~\cite{huffman2020}). For instance, for $N=4$ Dirac flavors, the bootstrap predicts $\nu^{-1}\approx 0.76$ while a wide variety of QMC methods give answers in the range $1.06\lesssim \nu^{-1}\lesssim 1.35$. For $N=8$, the bootstrap predicts $\nu^{-1}\approx 0.88$ while QMC predicts $1.0\lesssim\nu^{-1}\lesssim 1.3$. In both those cases, the previous QMC studies were sign-problem free and did not use SLAC fermions. This suggests the discrepancy for $\nu^{-1}$ in Table~I is due neither to the use of SLAC fermions, the presence of a sign problem, nor the choice of QMC method. Our work adds to the growing number of QMC studies of Ising GN criticality that challenge the existing bootstrap estimates for $\nu^{-1}$. Further and more accurate bootstrap studies of the GN universality classes are thus needed to resolve the discrepancy.

As a future direction, it would be interesting to apply the recently proposed adiabatic QMC algorithm~\citep{vaezi2020amelioration} to our model Hamiltonian and study the robustness of our results at considerably lower temperatures. Additionally, our study can be extended to other values of $N$, in particular the $N=1$ chiral Ising GN universality class which can be taken as an effective model of interacting Majorana surface states in the 3D topological superfluid $^3$He-B~\cite{mizushima2012,park2015}. Previous works on this universality class using the conformal bootstrap~\citep{Iliesiu2018,iliesiu2016}, fRG~\citep{PhysRevD.91.125003,Gies2017}, and perturbative RG~\citep{sonoda2011,grover2014,fei2016,PhysRevB.96.165133,PhysRevB.98.125109,PhysRevD.96.096010} have proposed that $\mathcal{N}=1$ spacetime supersymmetry emerges at the (2+1)D QCP. A numerical verification of this prediction would be of high value.

{\it Acknowledgments.}--- A.V. acknowledges useful discussions with Christian Mendl. S.M.T. and A.V. were supported by Iran Science Elites Federation (ISEF). J.M. was supported by NSERC Discovery Grants \#RGPIN-2020-06999 and \#RGPAS-2020-00064; the Canada Research Chair (CRC) Program; CIFAR; the Government of Alberta's Major Innovation Fund (MIF); the University of Alberta; the Tri-Agency New Frontiers in Research Fund (NFRF, Exploration Stream) and the Pacific Institute for the Mathematical Sciences (PIMS) Collaborative Research Group program. This research was enabled in part by support provided by Calcul Qu\'ebec (www.calculquebec.ca), Compute Ontario (www.computeontario.ca), WestGrid (www.westgrid.ca), and Compute Canada (www.computecanada.ca).

\bibliography{refs}

\begin{thebibliography}{50}
\bibitem [{CM2()}]{CM2}%
  \BibitemOpen
  \href@noop {} {}\bibinfo {note} {Furthermore, we chose the imaginary time
  step of $\Delta\tau = 0.05$ to keep the Trotter error well below our
  statistical error.}\BibitemShut {Stop}%
\end{thebibliography}

\pagebreak
\widetext
\clearpage


\section*{Supplementary material}

S. Mojtaba Tabatabaei,[1] Amir-Reza Negari,[1] Joseph Maciejko,[2] and Abolhassan Vaezi[1]

[1]*Department of Physics, Sharif University of Technology, Tehran 14588-89694, Iran*
[2]*Department of Physics & Theoretical Physics Institute (TPI),*
*University of Alberta, Edmonton, Alberta T6G 2E1, Canada*

In this supplemental material, we provide more details about the methods employed, and present our results from the mean-field calculation, finite-temperature quantum Monte Carlo (QMC) and also the finite-size corrections to the data collapse method for extracting the critical exponents.

### Mean-field calculations

Here we explain the mean-field approach to our lattice model with total Hamiltonian $H = H_0 + H_U$ where $H_0$ and $H_U$ are the non-interacting and interaction Hamiltonians given in Eqs. (2) and (4) of the main text, respectively. We start by rewriting $H$ in the mean-field approximation as:

$$H_{\text{MF}} = H_0 + Um/2 \sum_{\mathbf{i}} (n_{\mathbf{i}\uparrow} - n_{\mathbf{i}\downarrow}), \quad (\text{S1})$$

where $m = \sum_{\mathbf{i}} \langle n_{\mathbf{i}\uparrow} - n_{\mathbf{i}\downarrow}\rangle/N$ is the average order parameter in the FM phase. The mean-field Hamiltonian in Eq. (S1) is linear in terms of the fermion number operators, so it can be exactly diagonalized and the problem is then simply reduced to a self-consistent calculation of the order parameter $m$ for the ground-state expectation value of $\langle n_{\mathbf{i}\uparrow} - n_{\mathbf{i}\downarrow}\rangle$. Fig. S1 shows the results of mean-field calculations for system size $L = 31$. The increase of the FM order parameter above $U_c^{\text{MF}} \sim 3.68$, clearly shows that the system undergoes a SM-FM transition, although the critical interaction $U^{\text{MF}}$ that we obtained here is substantially smaller than the critical interaction that was obtained from exact QMC results.

### Determinant QMC algorithm

Here we discuss the implementation of the determinant quantum Monte Carlo (QMC) algorithm in more detail. By introducing a small parameter $\Delta\tau$ through $M\Delta\tau = \beta$, where $M \gg 1$ is the number of imaginary time steps, due to the Trotter-Suzuki formula the following expression can be considered as the density matrix:

$$\rho = \left(e^{-\Delta\tau(H_0+H_U)}\right)^M$$
$$= \left(e^{\frac{-\Delta\tau}{2}H_0} e^{-\Delta\tau H_U} e^{\frac{-\Delta\tau}{2}H_0}\right)^M + \mathcal{O}(\Delta\tau^3). \quad (\text{S2})$$

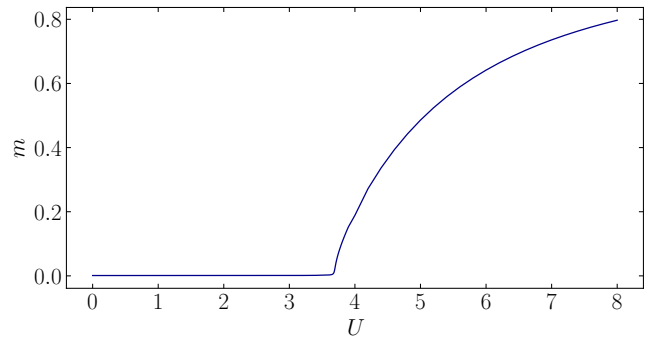

FIG. S1. Mean-field results for the FM order parameter, $m = \sum_{\mathbf{i}} \langle n_{\mathbf{i}\uparrow} - n_{\mathbf{i}\downarrow}\rangle/N$. A SM-FM transition in the mean-field approximation is seen at $U_c^{\text{MF}} \sim 3.68$. Here the system size is $L = 31$.

We found that a conservative value of $\Delta\tau = \frac{1}{20}$ can give rise to reliable results and a Trotter descritization error below our statistical error. The interaction Hamiltonian $H_U$ contains quartic terms which can be decoupled using a classical discrete field, a procedure known as the Hubbard–Stratonovich (HS) transformation. Generally, the average sign depends on the precise transformation, such that by choosing a suitable HS transformation we can mitigate the sign problem. The class of HS transformations we consider is:

$$e^{-\Delta\tau U(n_{\mathbf{i}\uparrow}-1/2)(n_{\mathbf{i}\downarrow}-1/2)} = \frac{1}{2}e^{-\frac{U\Delta\tau}{4}} \sum_{s_{\mathbf{i}}=\pm 1} e^{-2\lambda s_{\mathbf{i}}(\hat{\mathbf{s}}_{\mathbf{i}} \cdot \vec{l})}, \quad (\text{S3})$$

on a given lattice site $\mathbf{i}$ and for a given imaginary time step, where $\lambda$ is defined via $\cosh\lambda = \exp(\frac{U\Delta\tau}{2})$, $\hat{\mathbf{s}}_{\mathbf{i}}$ is the spin operator on site $\mathbf{i}$, and $\vec{l}$ is a unit vector in an arbitrary direction. The freedom to choose $\vec{l}$ means that we can decompose the interactions in different channels. While the HS transformation originally introduced by Hirsch [1] corresponds to the $s_z$ channel, we found that the average sign is significantly higher in the $s_x$ or $s_y$ channels. Indeed, with an $s_x$ decoupling, Fig. 1(a) in the main text shows that the average sign of our model at the QCP is not severe and we can approach the ground state by accessing reasonably low temperatures.

The reason for choosing $s_x$ (or $s_y$) decomposition channels is that decoupling the Hubbard interaction in the $s_x$ channel preserves the particle-hole symmetry in 1D systems (i.e., when $H_0 = p_x\sigma_x$) and guarantees the absence of the sign problem in 1D in the $s_x$ channel (which we

have verified numerically as well). However, the fact that our model is two-dimensional (and the presence of the $p_y\sigma_y$ term in $H_0$) breaks the aforementioned particle-hole symmetry and introduces the sign problem back into the model. However, in the $s_z$ channel, even in 1D the particle-hole symmetry is broken for any Hubbard-Stratonovich fields configurations. As a result, the average sign is more suppressed in the $s_z$ channel compared to the $s_x$ (or $s_y$) channels).

The average sign is decaying exponentially as a function of both the evolution time ($\beta$) and the system size ($N = L^2$). Fig. S2 displays the logarithm of the average sign of $L = 5, 7, \cdots, 21$ systems at $U = 7.275$ ($\beta = 2\Theta = 14$) as a function of the system size $N = L^2$. The plot is nearly linear for $L > 11$. Hence, the average sign indeed follows the expected exponential decay ($\langle\text{sign}\rangle \propto e^{-aN}$) for large enough $N$ values. We have also plotted in Fig. S3 the average sign at the same $U$ and for various $\beta = 2\Theta$ values and verify the $\langle\text{sign}\rangle \propto e^{-b\beta}$ behavior for $\beta > 4$.

Since the noise of QMC simulations is inversely proportional to the average sign, denoted $\langle\text{sign}\rangle$, the central limit theorem requires an increase in the Monte Carlo samplings and measurements by a factor of $1/\langle\text{sign}\rangle^2$ to achieve the same level of accuracy as in sign-problem-free models. For example, when $\langle\text{sign}\rangle = 1/5$, we need nearly 25 times more measurements. In this work, we have employed massively parallel QMC simulations on thousands of CPU cores to mitigate this problem. For each data point, we have taken up to several billion measurements to keep the statistical error below 0.2% and obtain highly accurate results. We have also considered an imaginary time step of $\Delta\tau = 0.05$ to keep the Trotter error well below our statistical error.

In this letter, we only consider odd system sizes ($L = 5, 7, 9, \ldots$). The reason for choosing odd values of $L$ in the SLAC fermion approach (in the current and past works) is that only for an odd number of sites are there states at exactly zero momentum ($p = 0$), and thus zero-energy states are achievable. For an even number of sites, the zero-momentum states (and thus the zero-energy modes) are absent (the lowest-allowed momenta are $p = \pm\pi/L$ in 1D) and therefore the finite-size effects are more pronounced. To minimize the finite-size effects and ensure a Dirac fermion zero mode, we focus on odd values of $L$.

To speed up calculations, instead of summing explicitly over Wick contractions, we use the following closed forms for 2-point and 4-point connected correlation functions of a fermion bilinear $\hat{A} = c^\dagger A c$:

$$\langle\hat{A}^2\rangle_s^C = \text{tr}\big[GA^2 - (GA)^2\big], \quad \text{(S4)}$$

$$\langle\hat{A}^4\rangle_s^C = \text{tr}\big[GA^4 - 4GAGA^3 - 3(GA^2)^2 + 12(GA)^2GA^2 - 6(GA)^4\big], \quad \text{(S5)}$$

where we define the fermion Green's function matrix as

$$G = I - \langle cc^\dagger\rangle_s. \quad \text{(S6)}$$

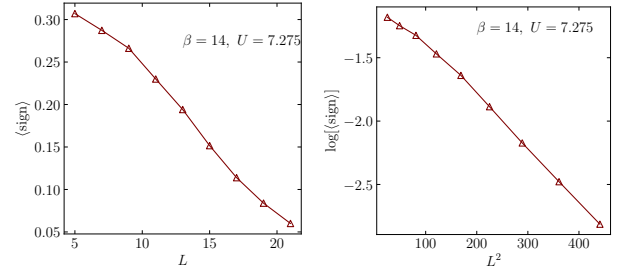

FIG. S2. Dependence of the average sign on the system size.

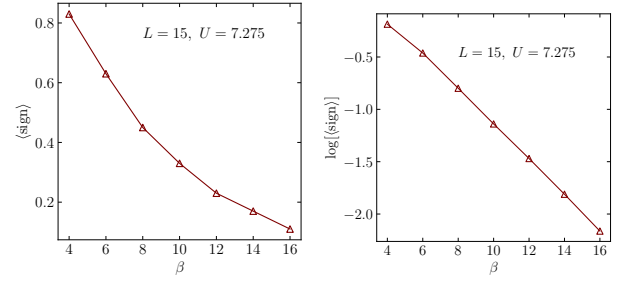

FIG. S3. Dependence of the average sign on the evolution time of the PQMC method ($\beta$).

Here $I$ is the identity matrix, $c$ is a vector of annihilation operators, and $A$ is a matrix which depends on the observable. In particular, this is useful for calculating the Binder ratio, which involves both 2-point and 4-point moments of the total spin operator, $\hat{S}_z$. In order to evaluate the Binder ratio, it is worthwhile mentioning that due to the time reversal symmetry of our microscopic model $\langle\hat{S}_z\rangle_s = \langle\hat{S}_z^3\rangle_s 0$ which implies the following identities between the cumulants and moments: $\langle\hat{S}_z^2\rangle_s = \langle\hat{S}_z^2\rangle_s^C$ and $\langle\hat{S}_z^4\rangle_s = \langle\hat{S}_z^4\rangle_s^C + 3\langle\hat{S}_z^2\rangle_s^2$. Therefore we have:

$$B = \frac{\langle\hat{S}_z^4\rangle}{\langle\hat{S}_z^2\rangle^2} = \frac{\sum_{ijkl}\langle s_{z,\mathbf{i}}s_{z,\mathbf{j}}s_{z,\mathbf{k}}s_{z,\mathbf{l}}\rangle}{\left(\sum_{\mathbf{ij}}\langle s_{z,\mathbf{i}}s_{z,\mathbf{j}}\rangle\right)^2}. \quad \text{(S7)}$$

where $\hat{S}_z$ is defined in terms of the fermion operators as:

$$\hat{S}_z = (c_{1\uparrow}^\dagger \cdots c_{L^2\uparrow}^\dagger c_{1\downarrow}^\dagger \cdots c_{L^2\downarrow}^\dagger)\begin{pmatrix} \mathbb{1} & 0 \\ 0 & -\mathbb{1} \end{pmatrix}\begin{pmatrix} c_{1\uparrow} \\ \vdots \\ c_{L^2\uparrow} \\ c_{1\downarrow} \\ \vdots \\ c_{L^2\downarrow} \end{pmatrix}, \quad \text{(S8)}$$

and 0 and $\mathbb{1}$ are the $L^2 \times L^2$ null and identity matrices, respectively. Without relations (S4-S5), we would have had to sum over four different indices and all possible Wick contractions of $\langle s_{z,\mathbf{i}}s_{z,\mathbf{j}}s_{z,\mathbf{k}}s_{z,\mathbf{l}}\rangle$ to find the value of $\langle\hat{S}_z^4\rangle_s$. Due to efficient matrix manipulation routines, our approach is significantly faster than performing this summation.

in the vicinity of the QCP:

$$S_{\mathbf{k}=0}(L,U,\beta) = L^{-(1+\eta_\phi)}(1+\alpha_1 L^{-\omega_1})$$
$$\times f_1\left((U-U_c)L^{1/\nu}, L/\beta\right), \quad (S9)$$

$$G_{\mathbf{k}=\mathbf{k}^*}(L,U,\beta) = L^{-\eta_\psi}(1+\alpha_2 L^{-\omega_2})$$
$$\times f_2\left((U-U_c)L^{1/\nu}, L/\beta\right). \quad (S10)$$

Here, we assumed the dynamical critical exponent $z=1$, which is a consequence of emergent Lorentz symmetry at the QCP. Based on this observation, we have two ways to make the scaling functions $f_1$ and $f_2$ depend on a single scaling variable. First, by considering the zero-temperature limit $\beta=\infty$, and second, at finite temperature but by considering isotropic scaling in spacetime, $\beta=L$. Then by setting all data points on a single plot with $(U-U_c)L^{1/\nu}$ on the horizontal axis, we can tune $U_c$, $\eta_\phi$, $\eta_\psi$, $\nu$, and the $\alpha$ and $\omega$ parameters such that all data points collapse on a single curve. With the finite-temperature method with $\beta=L$ scaling, large system sizes are typically required to reach the quantum critical regime and achieve sufficient accuracy for the critical exponents. For the system sizes we are able to reach, statistical noise in the finite-temperature data remains an obstacle to obtaining accurate exponents. By contrast, the projector QMC method allows us to reach the zero-temperature limit more effectively with limited system sizes. Thus, we focus on this method in the present work.

As we discussed in the main text, the fitting parameters $\alpha$ and $\omega$ in Eqs. (S9-S10) are not important for large system sizes (larger than $L=7$). In Fig. S4, we show the effect of including these correction terms. We can achieve a good data collapse even for system sizes as small as $L=5$ by choosing the following parameters for all four plots: $U_c=7.28$, $\nu^{-1}=1.19$, $\eta_\phi=0.31$, and $\eta_\psi=0.136$. We note that $\alpha$ and $\omega$ are not universal parameters and are chosen separately for each data set in Fig. S4, such that we can get the best data collapse.

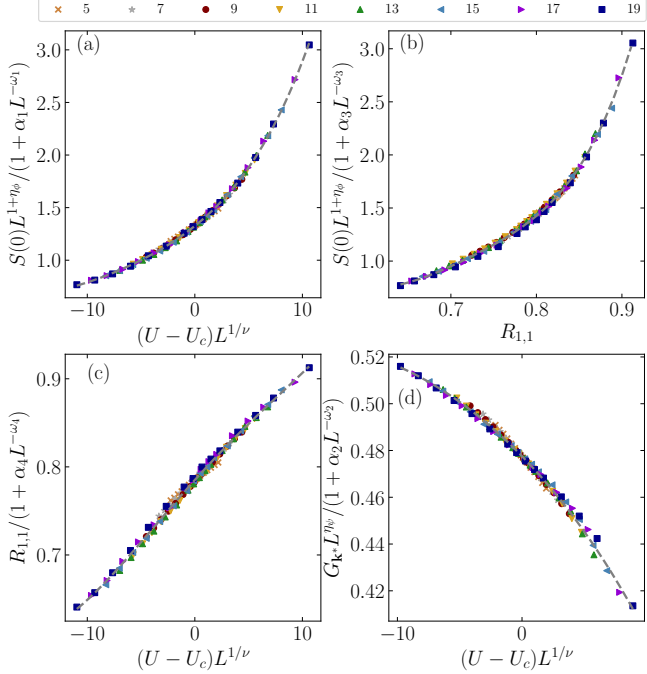

FIG. S4. Data collapse using the correction-to-scaling term given in Eqs. (S9-S10), for system sizes $5 \leqslant L \leqslant 19$. The fitting parameters $\alpha$ and $\omega$ are chosen separately for each data set such that we obtain the best data collapse. Here we used $U_c = 7.280$, $1/\nu = 1.19$, $\eta_\phi = 0.310$, and $\eta_\psi = 0.136$ for all plots. The non-universal fitting parameters are: $\alpha_1 = 1.04, \omega_1 = 2.22, \alpha_2 = 1.001, \omega_2 = 2.64, \alpha_3 = -88.25, \omega_3 = 3.96, \alpha_4 = 29.27$, and $\omega_4 = 3.96$.

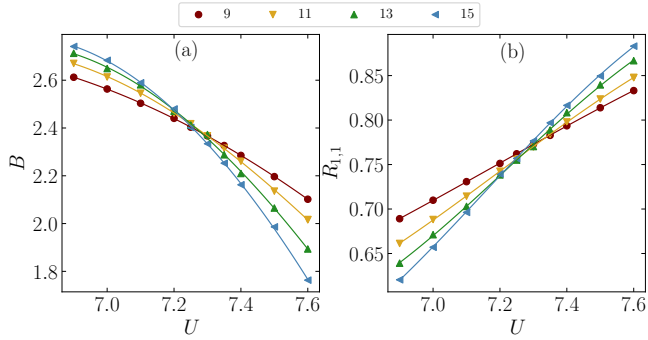

FIG. S5. (a) Binder ratio and (b) correlation ratio computed using finite-temperature determinant QMC with $\beta = L$ scaling, where $\beta$ is the inverse temperature and $9 \leqslant L \leqslant 15$ is the system size, denoted by different symbols.

### Finite-size scaling analysis

According to the scaling hypothesis, the ferromagnetic spin structure factor $S$ and the equal-time fermion Green's function $G$ should exhibit the following behavior

### Projector vs finite-temperature QMC

Although we have employed the projector QMC algorithm described in the main text as our main method, we have also performed finite-temperature determinant QMC calculations to corroborate our results. The finite-temperature determinant QMC method is based on discretizing the evolution operator in the computation of the finite-temperature statistical average $\langle O \rangle = \frac{\text{Tr}[e^{-\beta H}O]}{\text{Tr}[e^{-\beta H}]}$, using a symmetric Suzuki-Trotter decomposition scheme. Then, we can employ an HS transformation as previously discussed to handle the on-site Hubbard interaction. We consider $\beta = L$ scaling for $L = 5, 7, 9, 11, 13, 15, 17$. We extract the transition point using the Binder and correlation ratios. As Fig. S5 suggests, the transition lies

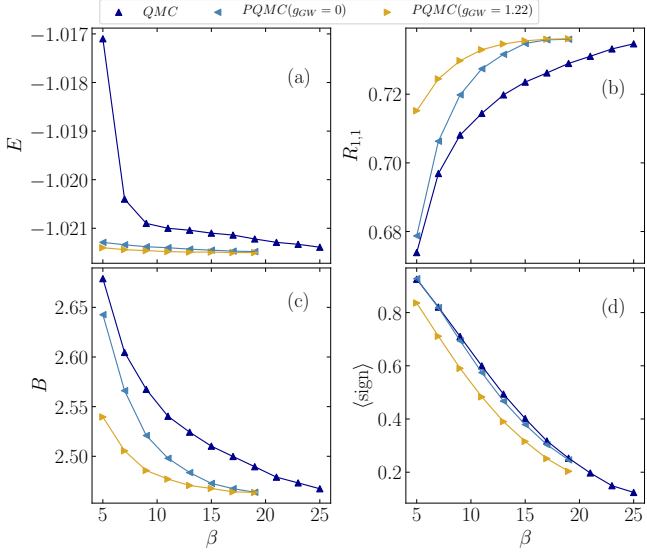

FIG. S6. Comparison between three different determinant QMC algorithms: finite-temperature QMC (dark blue), regular projector QMC (light blue) and projector QMC with a Gutzwiller-projected trial state with $g_{\rm GW} = 1.22$ (yellow), for $L = 11$ and $U = 7.1$. For the projector QMC plots, $\beta$ is defined as the value of $2\Theta$. (a) Energy density, (b) correlation ratio, (c) Binder ratio, (d) average sign. Projector QMC with $g_{\rm GW} \neq 0$ approaches the zero-temperature limit earlier than ordinary projector QMC and much earlier than finite-temperature QMC.

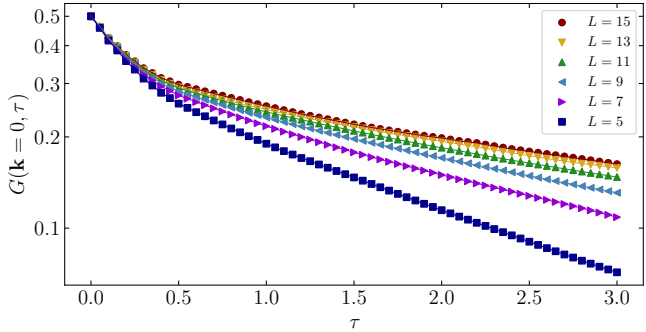

FIG. S7. Imaginary-time displaced Green's function $G(\mathbf{k} = 0, \tau)$ for various lattices sizes $L = 5, 7, 9, 11, 13, 15$. Other parameters are $U = 7.5$ and $2\Theta = 11$.

somewhere close to $U = 7.3$, which is consistent with the projector QMC results in the main text.

Finally, we compare the relative ability of projector QMC vs finite-temperature QMC to converge to the ground state, which is our main interest. In Fig. S6, we compute various quantities as a function of inverse temperature $\beta$ using three methods: finite-temperature QMC, regular projector QMC, and projector QMC with a Gutzwiller-projected trial wave function. For the projector QMC plots, we define $\beta \equiv 2\Theta$. We see that among all three methods, projector QMC with a Gutzwiller-projected trial state approaches ground-state properties the fastest. We can define an effective inverse temperature $\beta_{\rm eff}$ for a given $\Theta$ by comparing the value of physical quantities we obtain in projector QMC with those from finite-temperature QMC at $T = 1/\beta$. This reveals that $\beta_{\rm eff}(g_{\rm GW} \neq 0) \approx 2\Theta + 11 \pm 1$ and $\beta_{\rm eff}(g_{\rm GW} = 0) \approx 2\Theta + 7 \pm 1$. Thus, the projector QMC computations in the main text with projection time $2\Theta = 14$ translate effectively to finite-temperature QMC with $\beta_{\rm eff} = 25 \pm 1$. This is sufficient to accurately approximate zero-temperature critical properties.

**Extraction of single-particle excitation energy gap**

For imaginary-time displacement $\tau$, the single-particle excitation gap $\Delta_{\rm sp}$, leads to a behavior $G(\mathbf{k}, \tau) \equiv \langle c_{\mathbf{k}\uparrow}(\tau) c^{\dagger}_{\mathbf{k}\uparrow}(0) \rangle \propto e^{-\Delta_{\rm sp}(L,U)\tau}$ for the unequal-time fermion Green's function. It also implies the energy spectrum near the QCP is $(\mathbf{k}^2 + \Delta_{\rm sp}^2)^{1/2}$, which suggests $L\Delta_{\rm sp}(L, U)$ is a dimensionless quantity.

In order to extract the single-particle excitation energy gaps, we calculate the unequal-time fermion Green's function by implementing the approach proposed in Ref. [2] within our projector QMC calculations. Doing so, we calculate $G(\mathbf{k} = 0, \tau)$ for various values of $U$ and $L$. The excitation energy gap $\Delta_{\rm sp}$ is then obtained by fitting the tail of $G(\mathbf{k} = 0, \tau)$ to the form $e^{-\Delta_{\rm sp}\tau}$. We show a representative case for this calculation in Fig. S7.



\end{document}